\documentstyle[12pt]{article}
\input epsf
\newcommand{\const}{\mbox{{\sl const\/}}}
\newcommand{\beq}{\begin{equation}}
\newcommand{\eeq}{\end{equation}}

\newcommand{\dd}{D\hspace{-.65em}/}

\newcommand{\fr}[2]{{\textstyle {#1 \over #2}}}
\newcommand{\rf}[1]{(\ref{#1})}
\newcommand{\eq}[1]{Eq.\,(\ref{#1})}

\def\d{\partial}
\def\Om{\Omega}
\def\l{\lambda}

\def\P{\Phi}
\def\fr #1#2{\frac{#1}{#2}}

\def\op{operator}
\def\dop{Dirac operator}

\def\cn{condition}
\def\con{configuration}

\def\colvec{{\left(\begin{array}{c}1\\*0\end{array}\right)}}

\newcommand{\mod}{\rm{mod}\ }
\begin{document}
\begin{titlepage}
\begin{flushright}
HUTP-95/A036, NBI-HE-95-21, HEP-PH/9511240\\
Nov 1995\\
\end{flushright}
\vspace*{0cm}
\begin{center}
{\Large {\bf ${\bf Z}_2$ as the Topological Origin of\\[0.7ex]
B+L Violation in the Hot Electroweak Theory}}\\
\vspace*{0.7cm}
{\bf Minos Axenides}\footnote{e-mail:
{\tt axenides@nbivax.nbi.dk}}\\
\vspace{0.2cm}
{\em University of Crete, Department of Physics\\
 GR-71409 Iraklion-Crete, Greece}\\
\vspace*{0.4cm}
{\bf Andrei Johansen}\footnote{e-mail:
{\tt johansen@string.harvard.edu},
{\tt ajohansen@nbivax.nbi.dk}}\\
\vspace*{0.2cm}
{\em Harvard University, Lyman Laboratory of Physics\\
Cambridge, MA 02138, USA}\\
\vspace*{0.4cm}
{\bf Holger Bech Nielsen}\footnote{e-mail:
{\tt hbech@nbivax.nbi.dk}}\\
\vspace*{0.2cm}
{\em The Niels Bohr Institute,
University of Copenhagen\\ Blegdamsvej 17, DK-2100 Copenhagen \O, Denmark}\\
\vspace*{0.4cm}
{\bf Ola T\"{o}rnkvist}\footnote{Present address:
NASA/Fermilab Astrophysics Center, MS-209, P.O. Box 500,
Batavia,\\ \hspace*{1.5em} IL 60510, USA; e-mail:
{\tt olat@fnas13.fnal.gov}}\\
\vspace*{0.2cm}
{\em NORDITA\\
Blegdamsvej 17, DK-2100 Copenhagen \O, Denmark}
\end{center}

\begin{abstract}

The space of static finite-energy configurations in the electroweak
theory admits a ${\bf Z}_2$ topological structure. More precisely, we show that
this space contains two disconnected sectors of unstable
gauge-Higgs fields odd under a properly defined generalized parity.
This classification extends the description of baryon and lepton number
violating electroweak
processes to the symmetric phase of the theory.
Configurations with odd pure-gauge behaviour
at spatial infinity, such as the sphaleron, multisphalerons, electroweak
strings as well as an infinite surface of their equivalents,
have half-integer
Chern-Simons number and mediate $B+L$ violating processes in the early
universe.
Finite-energy configurations with even pure-gauge behaviour, such as the $S^*$
new sphaleron and electroweak strings, are
topologically equivalent to the vacuum and are irrelevant for $B+L$ violation.
We discuss the possible formation of $B+L$ violating quark-lepton
condensates in the symmetric high-temperature phase of the electroweak
theory.

\end{abstract}
\end{titlepage}
\newpage
\section{Introduction}
\setcounter{equation}{0}

Electroweak interactions are well described by models
with chiral fermions
and an $SU(2)\times U(1)$ gauge symmetry broken spontaneously to $U(1)$.
The interaction violates the conservation of baryon (B) and lepton (L)
quantum numbers through the chiral anomaly. As a consequence, the possibility
of generating the baryon-number asymmetry of the universe at the
electroweak phase transition has recently received increased attention
\cite{cohen}.

At zero temperature $B+L$ violation is negligible
and is a consequence of quantum tunneling between the
distinct vacua which constitute the topologically nontrivial periodic
structure of the $SU(N)$ Yang-Mills vacuum.
The tunneling
is induced by instantons in whose background the four-dimensional Dirac
operator
has normalizable zero modes \cite{hooft}.
The nontrivial structure is a result of
the existence of a homotopy group $\pi_3 (SU(N))={\bf Z}$.

At non-zero
temperature the density of instantons is believed to be suppressed
due to biscreening \cite{rob}.
Yet it has been argued \cite{peter} that for temperatures below the
$SU(2)\times U(1)$ phase transition temperature $T_{crit}$, the
thermal $B+L$ violating rates are dominated by ``sphaleron''
saddle point solutions \cite{manton,klink} to the electroweak field equations.
These solutions are static unstable finite-energy gauge-Higgs field
configurations with a Chern-Simons number $CS=1/2$ (modulo 1) relative to the
vacuum.
Their existence is a consequence of the periodic vacuum structure.
More precisely, a sphaleron configuration is the point of highest energy
($E_{sph}\approx
M_W/\alpha$) on the lowest-energy continuous
path of static field configurations that leads across
the potential energy barrier and interpolates between the
topologically distinct vacua with $CS=n$ and $CS=n+1$ ($n\in {\bf Z}$).
$B+L$ violating transitions across the sphaleron saddle point
are due to the crossing of fermion energy levels, which is a consequence of
the existence of normalizable zero-energy solutions to the three-dimensional
Dirac equation
in the sphaleron background \cite{kunz}.

The existence of the electroweak sphaleron is also a manifestation of the
nontrivial structure of the
infinite-dimensional space of static finite-energy configurations of the
electroweak theory.
Such a structure is hinted to by the existence of a non-contractible
loop in this space \cite{manton}.
The nontrivial structure was first demonstrated for the $SU(2)$ model
with an argument that depends crucially on the fact that
the symmetry is spontaneously broken.
In other words, the loop disappears in the symmetric phase of the model
($M_W\to 0$).
This construction therefore gives us no real information about the
structure of the configuration space in the symmetric phase.
The issue of this structure becomes perplexing when one tries
to relate, in the weak coupling limit, the $CS=1/2$ sphaleron $S$ to
the newly found $CS=0$
axisymmetric $S^*$ ``new sphaleron'' solution \cite{kl}
which can be interpreted as a bound state of a sphaleron and an
antisphaleron. More generally the structure should account
for the existence of recently discovered multisphaleron
bound-state solutions
\cite{Kleihaus}
with $CS= n/2$, $n\in{\bf Z}$.
What remains a puzzle is the relation of the $S$, $S^*$, and multisphaleron
configurations to the "bifurcated"
saddle points with $CS=1/2\pm \epsilon({\rm m}_H)$ found by Yaffe for
large Higgs masses \cite{yaffe}.

It is expected that, in the presence of large thermal
fluctuations of the gauge and Higgs fields in the hot electroweak plasma,
sphaleron deformations
with energy equal to or higher than that of the sphaleron will
contribute to rapid baryon and lepton-violating transition rates.
Their presence is especially likely to dominate the symmetric
high-temperature phase of the electroweak theory ($T\geq E_{sph}\sim
M_W/\alpha$). In fact scaling arguments indicate unsuppressed $B+L$
violating transition rates ($\Gamma\propto T^4$) in this r\'{e}gime
\cite{temp}.
Recent computer simulations of hot electroweak sphaleron transitions
in the symmetric phase \cite{ambjorn}
corroborate to this physical picture.

In two previous publications \cite{topclass,ajm} it was observed that the $B+L$
violating properties of the electroweak $SU(2)$ sphaleron follow
directly from its odd-parity gauge-field properties.
More precisely, as a consequence of the existence of ${\bf Z}_2$
homotopic groups of maps $[S^2/{\bf Z}_2 , SU(2)/{\bf Z}_2 ]={\bf Z}_2$
and $[S^3/{\bf Z}_2 , SU(2)/{\bf Z}_2 ]={\bf Z} \times {\bf Z}_2$,
the Chern-Simons functional restricted to odd-parity gauge fields
is a topological charge.
It takes values in ${\bf Z}$ or in ${\bf Z}+1/2$ depending on whether
the fields at spatial infinity have even-parity or odd-parity pure-gauge
behaviour respectively.
Hence the existence of the $SU(2)$ sphaleron as well as of an infinite
surface of its homotopically  equivalent deformations, such as loops of
electroweak $W$ string \cite{Wstring,embedded},
has a novel group-theoretic explanation.

An important implication from such a topological index for the
spectrum of the three-dimensional Dirac operator is that the number of
zero modes modulo 2 is a topological invariant.
Indeed in the presence of an odd-parity external gauge field the number
of fermionic zero modes is either $0~(\mod 2)$ or $1~(\mod 2)$
for fields with
respectively even or odd pure-gauge behaviour at spatial infinity.
We remark that the existence of ${\bf Z}_2$ homotopy groups is
the property only of Lie groups whose centers have ${\bf Z}_2$
as a subgroup. These are $SU(2N)$, $SO(2N)$ and $E_7$ \cite{ajm}.
For gauge groups which do not have ${\bf Z}_2$ in their center the structure is
more complicated.

In the present paper we demonstrate that the ${\bf Z}_2$ topological structure
in the
configuration space of a general $SU(2)\times U(1)$ model implies the
existence of two topologically disconnected classes of
gauge-field configurations which are odd under
a generalized parity transformation,
defined as the composition of an ordinary parity transformation
and a gauge transformation. Eigenvalues under generalized parity are
gauge invariant. The part of configuration space associated with the ${\bf
Z}_2$
structure is infinitely large.

Furthermore we show that the ${\bf Z}_2$ structure
accounts for the new $S^*$ ``sphaleron'' with $B+L$ preserving
properties \cite{kl} as well as other $B+L$ preserving
configurations constructed
from $W$-string loops and $Z$-string segments.
It moreover provides us with sufficient conditions for the construction
of infinitely many sphaleron deformations with $B+L$ violating properties,
such as $W$-string loops, properly twisted segments of
$Z$ string connecting a monopole with an
antimonopole, and multisphaleron configurations with $CS=n+1/2$,
$n\in {\bf Z}$.
The plethora of examples of both kinds suggests
that a hot electroweak plasma at
sufficiently high temperature can be described qualitatively
as a ``two-component
fluid'' composed of two types of localized excitations corresponding to
configurations which permit or prohibit $B+L$ violation. The density of
both types of excitations is nonzero in the symmetric phase, while the density
of the $B+L$ violating component
decreases rapidly to zero for $T<T_{crit}$.
We speculate about the possible existence in the symmetric phase
of quark-lepton condensates in the presence of such a non-zero density of $B+L$
violating gauge-Higgs field configurations.

The rest of the paper is organized as follows.
In section 2 we review a topological classification for
odd-parity $SU(2)$ gauge fields with pure-gauge behaviour at spatial infinity.
In section 3 we extend this approach to $SU(2)$ gauge fields which are odd
under a generalized parity transformation.
We also generalize our approach to the $SU(2)\times U(1)$
theory.
In section 4 we compute the Chern-Simons number of gauge-field configurations
which are odd under generalized parity.
In section 5 we construct explicit examples of sphaleron
deformations using twisted loops of electroweak
$W$ string as well as $Z$-string \cite{Zstring} segments connecting a
monopole with an antimonopole (Nambu strings) \cite{nambu}.
Single twisted loops of $Z$ string
are shown to lead to inconsistent field configurations and can exist only
when linked with another vortex string. The $B+L$
preserving properties of the new $S^*$ ``sphaleron'' are established,
as well as the $B+L$ violating properties of
multisphaleron configurations with $CS=n+1/2$, $n\in{\bf Z}$.
The topological ${\bf Z}_2$ structure implies the existence also of $B+L$
preserving multisphaleron solutions with $CS=n\in {\bf Z}$.
We conclude with a discussion of the possible formation of fermionic
condensates in the high-temperature electroweak symmetric phase in the
presence of a non-zero density of localized excitations which correspond to
configurations that admit an odd number of fermion zero
modes.

\section{An SU(2) Topological Classification}
\setcounter{equation}{0}

We start with the observation that the static
sphaleron configuration has an odd-parity gauge field everywhere in
space.
While this occurs in a particular gauge it will prove sufficient for our
purposes, because physical properties such as its energy and the Dirac spectrum
in its
background are gauge independent.
By imposing parity oddness on all possible deformations (they may or may not
be solutions) we find two topologically distinct sectors of
configurations that depend on the (even-odd) parity property of the
group-valued function that determines their pure-gauge behaviour at spatial
infinity.

Let us recall some properties of the sphaleron configuration.
Its gauge field reads
\beq
A_i = v(r) \frac{\epsilon_{ijk} x_j \tau_k}{r^2}
= -iv(r) \partial_i U_{sph} \; U_{sph}^{-1} ,
\label{spha}
\eeq
where $\tau_k$ are the Pauli matrices, $r^2=x_j x_j$, Roman indices run over
the
three space components, and
\beq
\label{sphaU}
U_{sph} = \frac{ix_k \tau_k}{r}.
\eeq
The profile function $v(r) \to 1$ as $r\to \infty$
so that this configuration is a pure gauge at infinity.

The Chern-Simons functional is defined to be
\beq
CS[A] = \frac{1}{8\pi^2} \int_{D^3} {\rm Tr}(A dA - \frac{2i}{3} A^3)
\eeq
and is gauge dependent. In the particular gauge defined by Eqs.\,\rf{spha} and
\rf{sphaU} the
Chern-Simons functional of the sphaleron configuration evaluates to zero.
This fact can be checked almost without calculation due to
the observation that this gauge-field configuration is odd under parity.
However, the Chern-Simons {\it number\/} ($CS$) of the sphaleron is correctly
defined as
the gauge-independent difference in the value of the Chern-Simons functional
when applied
to the sphaleron
configuration and to the vacuum. In order to make this comparison,
we must transform to a gauge in which the gauge
 fields decrease more rapidly than $1/r$ at infinity. In such a gauge, the
vacuum has zero $CS$ number and the sphaleron has $CS=1/2$, as we show below.
 The gauge transformation is accomplished by an $SU(2)$ group element $U'$
which is smooth everywhere and coincides with $U_{sph}$ at infinity.
Since $\pi_2 (SU(2)) =0$ we know that such a field $U'$ does exist.
The Chern-Simons functional transforms gauge covariantly according to the
relation
\beq
{\rm CS} [A'] ={\rm CS} [A] - S_{WZW} [U'] ,
\eeq
where
\beq
A'_k = U' A_k (U')^{-1} -i \partial_k U' \; (U')^{-1} .
\eeq
and $S_{WZW}$ stands for the Wess-Zumino-Witten functional defined as follows
\beq
S_{WZW}[U'] = \frac{1}{24\pi^2} \int_{D^3} {\rm Tr} \;(dU'\; U'^{-1})^3 .
\eeq
This term actually depends only on the behaviour of the field $U'$
at the boundary $S^2$ of the disk $D^3$ and is equal to 1/2 for any
field $U'$ with the asymptotic property $U'\to U_{sph}$.
For example we can take
\beq
U' = \exp \frac{i\pi}{2} \frac{\tau_i x_i}{\sqrt{x^2 + \rho^2}}
\eeq
and check by an explicit calculation that $S_{WZW} [U'] =-1/2.$
In turn ${\rm CS} [A] =0$ since the field $A$ is odd under parity, i.e.
$A_k(-x) = -A_k(x).$
Thus we conclude that ${\rm CS} [A'] =1/2$, which defines the $CS$ number of
the
sphaleron.
It is now clear that the same value of the Chern-Simons number
corresponds to all odd-parity configurations which behave as $U_{sph}$
at infinity.

We see that the odd parity of the spherical
sphaleron is essential in order for the Chern-Simons number of the
sphaleron to be exactly $1/2$. However, this is not sufficient by itself.
Actually, as we shall see below, the half-integer value of the $CS$ number also
depends crucially on the odd parity of $U'$ at infinity.
In a previous publication \cite{topclass} a connection was established
between the Chern-Simons number of gauge fields and the parity
property of their pure-gauge behaviour (i.e. the $U'$ field itself)
at infinity.
We argued there that a restriction to odd-parity
gauge-field configurations allows us to introduce a useful
topological classification among these fields.

For gauge fields which are pure gauge
at infinity (i.e. on the boundary of a
3-dimensional ball, which is $S^2$)
\beq
A_i = -i(\partial_i U)\; U^{-1},
\eeq
where $U$ belongs to the $SU(2)$ group.
The restriction of this field $U$ to the boundary $S^2$
of the 3-dimensional ball is a map of $S^2$ into $SU(2) .$
The homotopic group $\pi_2 (SU(2))$ is trivial and hence
all such configurations in the 3-dimensional ball are contractible to
unity.
We now restrict ourselves to the space of 3-dimensional
odd-parity gauge fields.
We shall argue that
there exists a relevant non-trivial homotopic classification in this space of
gauge fields.
Indeed let us consider an odd-parity configuration
\beq
A_i (-x) = - A_i (x).
\eeq
On the $S^2$ boundary this is a pure gauge so that
\beq
(\partial_i U) (-x) U^{-1} (-x) = -
(\partial_i U) (x) U^{-1} (x) ,
\eeq
or equivalently
\beq
\label{newder}
\d_i\!\left[U^{-1}(-x)U(x)\right] = 0 .
\eeq
{}From this it follows that the matrix $V=U^{-1}(-x)U(x)$ is a non-degenerate
constant $SU(2)$
element, and consequently the equation
\beq
U(x) = U(-x) V
\eeq
is valid for any point $x$ on  $S^2 .$
{}From this equation, combined with that obtained by changing $x \to -x$, we
find
\beq
V^2 =1 .
\eeq
A short inspection now shows that the matrix $V$ should belong to
the center of $SU(2)$. In equations
\beq
V= \pm 1 .
\eeq
Thus we get two different classes of odd-parity gauge fields:
Those with odd $U$ and those with even $U .$

This conclusion reflects a non-triviality of the homotopy
group of maps from the projective sphere
${\rm {\bf R}}P^2 = S^2 /{\bf Z}_2$ (where ${\bf Z}_2$ is a group of
parity transformations with respect to some point in 3-dimensional
space) to the group
$SO(3) = SU(2) /{\bf Z}_2$ (where ${\bf Z}_2$ is the center of $SU(2)$).
In short $[{\bf R} P^2 ,SO(3)] = {\bf Z}_2 $.
The consideration above shows that
the odd-parity gauge fields split into two topologically
disconnected equivalence classes.
In other words it is not possible to get from one to the other
continuously through odd-parity gauge-field configurations.

Actually the classification of these gauge fields is more complicated.
Instead of restricting to the sphere $S^2$, we must take into
consideration the set of gauge-field configurations $A_i(x)$
which are odd under parity in the entire 3-dimensional
ball. The relevant homotopy group is then
$[{\bf R}P^3, SO(3)] = {\bf Z}\times{\bf Z}_2.$ \footnote{Notice that this
statement is not quite rigorous from the mathematical point of view. Actually
as it is shown in Ref. [13] the correct formulation
is $[{\rm R} P^3 , SO(3)]={\bf Z}$ where ${\bf Z}$
is doubly covered. Such a double covering is induced by the map
$SU(2)\to
SO(3)=SU(2)/{\bf Z}_2$.}
The factor ${\bf Z}_2$  can be identified with the above
classification in terms of odd $U$ or even $U$.
Each of these two classes of odd-parity fields thus possesses its own
${\bf Z}$ structure of topologically disconnected components. For
fields with even $U$ this structure is directly related to
the periodic structure of the Yang-Mills vacuum. More precisely, each component
contains one of the vacua
 $A_i^{n}= -i(\partial_i U_n)
U_n^{-1}$, where the group element $U_n$ is given by the even-parity
(at infinity) group elements \cite{ryder}
\beq
U_n = \exp (in\pi \frac{\tau_i x_i}{\sqrt{x^2 + \rho^2}}) .
\eeq
Here $\rho$ is a constant parameter and $n$ is the integer Chern-Simons number
of a vacuum configuration. Because there is no additional structure, it follows
that any configuration with an even $U$ field at infinity is continuously
connected to one of these vacua.

As we saw above, the sphaleron gauge field is (in an appropriate
gauge) odd under parity and has an odd $U$ (see \eq{sphaU}) and
hence it is disconnected from the configurations with even $U$.
The ${\bf Z}$ structure of the fields with odd $U$ at infinity
implies
the existence of an infinite sequence of topologically distinct
sphaleron solutions
associated
with saddle points on the energy barriers separating neighboring vacua.
Because there is no additional structure, any configuration with an odd
$U$ field at infinity will be continuously connected to one of these
sphalerons.

A common feature of all odd-parity, even-$U$ configurations is that they
have an integer-valued Chern-Simons number.
Indeed, in analogy with the case of the sphaleron we can make a nonsingular
gauge rotation which removes the gauge field at infinity.
The Wess-Zumino-Witten functional would then give us the Chern-Simons
number of the gauge-field configuration.
It is easy to see that the value of $S_{WZW}[U]$ is
invariant under even-parity
smooth deformations of the even $U$ field at infinity.
Indeed a variation of the Wess-Zumino-Witten functional reads
\beq
\label{WZWvar}
\delta S_{WZW}[U] = \frac{1}{8\pi^2} \int_{D^3} d {\rm Tr}
((U^{-1} \delta U)(U^{-1}dU)^2 )= \frac{1}{8\pi^2} \int_{S^2} {\rm Tr}
(U^{-1}\delta U)(U^{-1}dU)^2 .
\eeq
Since the variation of the group element on the surface $\partial D^3 =S^2$ is
even under parity and its value depends only on the values of the fields
at the boundary we immediately conclude that the present variation of the
Wess-Zumino-Witten functional equals zero.
On the other hand, let us consider
a product of even-parity (at infinity) group elements $U_1$ and $U_2$.
It is easy to check that
\beq
\label{hmorph}
S_{WZW} [U_1 U_2] = S_{WZW} [U_1] +S_{WZW} [U_2]
+ \frac{1}{8\pi^2} \int_{D^3} d\;{\rm Tr}\,((U^{-1}_1 dU_1 ) (dU_2 \;
U_2^{-1})).
\eeq
The third term in the right-hand side of this equation equals zero
due to the odd parity of the integrand at infinity.
Hence we see that the Wess-Zumino-Witten functional acts as a homomorphism
from the group of maps $U$ to a discrete
subgroup of the group of real numbers which is obviously
isomorphic to ${\bf Z} .$
As we argued before, any even-$U$ (at infinity) group element
is contractible to a vacuum.
On the other hand, as it is well known that this vacuum can have any integer
value of the
Chern-Simons number, we conclude that even-parity $U$ fields are indeed
classified by ${\bf Z}.$

Thus all odd-parity even-$U$ gauge fields split into
an infinite set of disconnected equivalence classes which are
labeled by integer values of their Chern-Simons numbers.

Let us now consider odd-parity odd-$U$ gauge fields.
A similar argument shows that the value of the Chern-Simons functional
is a topological invariant while the Wess-Zumino-Witten functional
maps the odd-parity $U$ fields to a discrete subgroup of the group of
real numbers according to \eq{hmorph}.
On the other hand a product of two odd-parity group elements
$U_1$ and $U_2$ is even under parity.
By taking into account that the sphaleron has
$S_{WZW} [U] =-1/2$ we conclude
that the odd-parity odd-$U$ gauge fields have half-integer
values of the Wess-Zumino-Witten functional and hence are classified
by $n+1/2$ ($n\in {\bf Z}$),
while these equivalence classes are themselves topologically
disconnected from each other for different values of $n.$

Thus we see that the Chern-Simons number plays the role of
a topological charge: It takes values in ${\bf Z}$
for even-$U$ and in ${\bf Z} + 1/2$ for odd-$U$ (at infinity) fields
respectively.

An immediate implication is that the Chern-Simons number acts as a topological
index for
the spectrum of the 3-dimensional \dop{}.
Let us consider a \dop{} $\dd = \sigma_i
(\partial_i -iA_i)$ in an external odd-parity gauge field $A_i .$
It is easy to see that the
non-zero eigenvalues appear in pairs ($\lambda, -\lambda$).
Indeed if $\psi(x)$ is a wave function corresponding to an eigenvalue
$\lambda$ then $\psi(-x)$ is an eigenfunction corresponding to an
eigenvalue $-\lambda .$
Hence, when the external field varies continuously, the number of
zero modes of the \dop \ is invariant modulo 2.
For the sphaleron background this topological invariant is equal to one
while for the vacuum its value is zero.

Since odd-$U$ configurations are continuously connected to a sphaleron,
and even-$U$ configurations to a vacuum, the invariance modulo 2
of the number of zero modes implies that the number of fermionic zero modes
in
an odd-parity background gauge field is 0 mod 2 for even-$U$ and 1 mod 2 for
odd-$U$
configurations.

\section{A Generalized Parity Symmetry}
\setcounter{equation}{0}

In this section we extend our topological classification to gauge fields which
are non-odd under a parity transformation.

Let us consider a semi-simple group $G$.
We define generalized parity to be the parity up to a gauge transformation.
Let us consider gauge fields odd under generalized parity, i.e.\ fields that
satisfy the equation
\begin{eqnarray}
A_i (-x)= -A_i^S (x) &\equiv& -S(x) (i \partial_i + A_i (x))S^{-1}(x)
\nonumber\\*
&=&-S(x) A_i (x) S^{-1}(x) + i\partial_i S(x) S^{-1}(x) ,
\label{AparG}
\end{eqnarray}
for some $S(x) \in G.$
As a trivial example, consider an ordinary odd-parity gauge field which has
been
transformed with an arbitrary parameter of the gauge group. The transformed
field is then odd under generalized parity. On the other hand, if
we pick a gauge field which is odd under generalized
parity, one can show that it is still odd under generalized parity after an
arbitrary gauge transformation. Therefore, generalized parity symmetry is
a gauge-invariant property. However, as we shall show below, the set of gauge
fields odd under generalized parity is larger than the set of gauge fields
odd under
ordinary parity. For the continued discussion, we consider the gauge groups
$SU(2)$ and $SU(2)\times U(1)$ separately.

\subsection{The Case of SU(2) Generalized Parity}

Consider a field $A_i(x)$ which is odd under the generalized parity
transformation
defined by $S(x)\in SU(2)$.
Then
\beq
\label{ftrans}
F_i (-x) = S(x) F_i (x) S^{-1} (x) ,
\label{FparG}
\eeq
where
\beq
F_i = \frac{1}{2} \epsilon_{ijk} F_{jk} ,\;\;\; i=1,2,3,
\eeq
and $F_{jk}$ is the strength of the gauge field.
Combining \eq{FparG} with the same equation for
$x\to -x$, one obtains
\beq
F_i (x) = \Om (x) F_i (x) \Om^{-1}(x) ,
\label{triv}
\eeq
where
\beq
\Om (x) = S(-x) S(x).
\eeq
{}From \eq{triv} one can see that $\Om$ is diagonalized simultaneously
with three components $F_i .$
There are two possibilities:
\begin{itemize}
\item[i)] $\Om = \pm 1 ,$
\item[ii)] $F_i$ gets factorized as follows
\beq
F_i (x) = f_i (x) \cdot i\P (x) ,
\label{factF}
\eeq
where $f_i$ are scalar functions, $\P (x)$ belongs to the
$SU(2)$ algebra, and
\beq
\Om (x) = \exp (i\l (x) \P (x)) .
\label{factO}
\eeq
\end{itemize}
Here $\l (x)$ is a scalar function.

In the second case the gauge field $A_i$ belongs to an abelian subgroup of
$SU(2)$, and hence can be
taken to be proportional to the $\tau_3$ generator of the $SU(2)$ algebra.
In what follows we focus on the first case, which more interesting
because it corresponds to non-abelian gauge fields.

Let us find a necessary \cn{} for the gauge field $A_i$ that
obeys \eq{AparG} not to be gauge
equivalent to a gauge field $\tilde{A}_i$ which is odd under the ordinary
parity
transformation (i.e. $\tilde{A}_i(-x) = -\tilde{A}_i(x))$.
In terms of the field strength, such a reducibility would imply that we
can find a gauge transformation $T$ such that
\beq
\tilde{F}_i (x) = T(x) F_i (x) T^{-1} (x)
\eeq
where $\tilde{F}$ is even under the usual parity transformation,
\beq
\tilde{F}_i (-x) = \tilde{F}_i (x) .
\eeq
{}From these equations and \eq{ftrans} it follows that
\beq
T(-x) S(x) T^{-1} (x) = \pm 1
\eeq
or, equivalently,
\beq
S(x) = \pm T^{-1} (-x) T(x) .
\eeq
Combining this equation with that obtained by taking  $x\to -x$ we easily get
\beq
\Om (x) =1.
\eeq

In this paper we concentrate on the simplest case of gauge fields which
are invariant under a generalized parity transformations with
$S(x)\equiv B=\const$. Therefore let us
apply the above analysis to this special case. One easily gets from the above
equations that
$$B^2 =1,$$
and hence (since $B\in SU(2)$)
\beq
S(x)=B=\pm 1,
\eeq
which implies that the gauge field $A_i(x)$ is odd under an {\it ordinary}
parity
transformation.
Therefore we see that there exist gauge
field configurations invariant under generalized parity with a constant $S$
which are not reducible to gauge configurations which are invariant
under an ordinary parity transformation.
Thus the notion of a generalized parity is not empty.
In section 5 we shall give explicit examples of such configurations.

We shall now proceed by showing that, for $SU(2)$ gauge fields with odd
generalized
parity,
there exists a non-trivial homotopic classification in terms of the parity
properties
 of the $SU(2)$-valued function $U(x)$ which determines the pure gauge behavior
\beq
\label{su2inf}
A_i (x) \sim -i(\partial_i U U^{-1})(x)
\eeq
at infinity.
By using eq.(3.1)
at spatial infinity we find that
\beq
(\partial_i U U^{-1})(-x) = - B(\partial_i U U^{-1})(x) B^{-1} ,
\eeq
and therefore
\beq
\partial_i (U^{-1}(-x)BU(x))=0.
\eeq
At infinity we must therefore have that
\beq
BU(x) = U(-x) V,
\label{BUV}
\eeq
where $V \in SU(2)$ is a constant matrix.
Using this equation and its parity conjugate one easily gets
\beq
B^2 U(x) = U(x) V^2 ,
\eeq
and thus
\beq
U(x)^{-1} B^2 U(x) = V^2 \in SU(2) .
\label{squaredB}
\eeq

Let us analyze the different cases for the matrix $B^2$. Let us first assume
that
$B^2\neq \pm 1$.
Then from \eq{squaredB} we get
\beq
U(x)^{-1} B U(x) = \pm V .
\label{unsquB}
\eeq
Combining \eq{unsquB} with \eq{BUV} we obtain
\beq
U(-x) = \pm U(x) ,
\eeq
from which it can be checked that $B^2=\pm 1$ in contradiction of our
assumption.
Therefore we conclude that $B^2 =V^2 \in {\bf Z}_2$. Then there are two
possible cases.
\begin{itemize}
\item[1.]
If $B^2 = 1$ then
$B=\pm 1$. Then $U(-x) = \pm U(x)$, which is the case of odd or even $U$
found already in section 2.
\item[2.]
If $B^2 =-1$ then $B$ belongs to the coajoint orbit ${\cal B} =
SU(2)/U(1) ,$ i.e.
\beq
B \in {\cal B} = \{ i S^{\dag} \sigma_3 S | S\in SU(2) \} .
\eeq
\end{itemize}
This orbit is isomorphic to an $S^2$ sphere.

Thus we conclude that there exist the following two classes:
\begin{eqnarray}
\label{su2classes}
{\bf 1.}&& U(x) = \pm U(-x) ,\\*[1ex]
{\bf 2.}&&\tau U(x) = \pm U(-x) \tau' ,\;\;\; i\tau, \; i\tau'\in {\cal B} .
\label{newclass}
\end{eqnarray}
We see that the condition of class 2
can be satisfied by a constant matrix $U(x)= \const.$
This class is then topologically trivial, because
it intersects with the class of $U$-even configurations $U(x)=U(-x)$
which contains the vacuum.
Thus the topologically non-trivial configurations satisfy $U(-x) = -U(x)$ and
belong to class 1,
which has been analyzed in the previous section
and in Ref.~\cite{topclass}.

\subsection{The Case of SU(2)$\times$U(1) Generalized Parity}
Let us now consider the case of an $SU(2)\times U(1)$ gauge theory.
We focus on the gauge fields which obey the following condition
(a generalized parity)
\beq
A(-x) = - B A(x) B^{\dag},
\label{constU}
\eeq
where $B$ is a constant matrix
belonging to $SU(2)/{\bf Z}_2,$ ${\bf Z}_2=\pm 1$ is the center of $SU(2),$
and $B\neq \pm 1$.

At infinity we have
\beq
A_i (x) = -i(\partial_i U U^{-1})(x),
\eeq
where $U(x) \in U(2) .$
Hence
\beq
(\partial_i U U^{-1})(-x) = - B(\partial_i U U^{-1})(x) B^{\dag}.
\eeq
Therefore we conclude that
\beq
BU(x) = U(-x) V,
\label{BUVU}
\eeq
where $V \in U(2)$ is constant matrix.

Using this equation twice we get
\beq
B^2 U(x) = U(x) V^2 ,
\eeq
and thus
\beq
U(x)^{-1} B^2 U(x) = V^2 \in SU(2) .
\label{squU}
\eeq
We may classify the different cases for the matrix $B^2$:

{\bf 1.} $B^2 \in {\bf Z}_2$, i.e. $B^2 =\pm 1 .$
Then $V^2 \in {\bf Z}_2$ and $B^2 = V^2 .$
Since $B$ does not belong to ${\bf Z}_2$ (the center of $SU(2)$) we have two
subcases

(1a). $B = i S^{\dag} \sigma_3 S$, $V= \pm i$ or $V=S'^{\dag} i\sigma_3 S' ,$
where $S, S' \in SU(2) .$

(1b). $B = S^{\dag} \sigma_3 S$, $V= \pm 1$ or $V=S'^{\dag} \sigma_3 S' ,$
where $S, S' \in SU(2) .$

{\bf 2. $B^2 \neq \pm 1 .$}
Then from \eq{squU} we get
\beq
U(x)^{-1} B U(x) = \pm V .
\label{unsquU}
\eeq
Substituting \eq{unsquU} into \eq{BUVU} we come to the \cn \
$U(-x) = \pm U(x) ,$ i.e. to the old case of odd and even $U(x) .$

As a result we have the following classification of pure-gauge
behaviour of the gauge fields odd under a generalized parity
defined in \eq{constU} (let us denote this
subspace of the space of the gauge fields as ${\cal A}^P$) :
\begin{eqnarray}
\label{Uone}
{\bf 1.}&& \tau U(x) = \pm U(-x) ,\;\;\; i\tau \in {\cal B} ,\\*[1ex]
\label{Utwo}
{\bf 2.}&&\tau U(x) = \pm U(-x) \tau' ,\;\;\; i\tau, \; i\tau'\in {\cal B},
\\*[1ex]
\label{Uthree}
{\bf 3.}&&U(x) = \pm U(-x)\tau ,\;\;\; i\tau \in {\cal B},\\*[1ex]
\label{Ufour}
{\bf 4.}&&U(x) = \pm U(-x) .
\end{eqnarray}
Now we can look at each of the above four classes.
Class 4 (\eq{Ufour}) corresponds to gauge fields odd under the ordinary
parity transformation.
For the case of an $SU(2)$ gauge group such fields have been
classified in Ref. \cite{topclass}.
There it was shown that the gauge fields with
pure-gauge behaviour at infinity corresponding to an
even-parity group element are topologically trivial and have integer
Chern-Simons number.
The gauge fields with odd-parity pure-gauge behaviour at infinity
are topologically non-trivial and have half-integer Chern-Simons number.
Here we see that an extension of that analysis to the case of
the $SU(2)\times U(1)$ group is straightforward; the classification
of odd-parity gauge fields is the same as with the case of the $SU(2)$ group.

Let us now look at the other cases.
Class 2. (\eq{Utwo}) can be satisfied by a constant matrix $U(x)= \const.$
Hence it is topologically trivial since it intersects with the
class of fields obeying the \cn \ $U(-x) =U(x)$ such as the trivial
vacuum.

Classes 1. and 3. (eqs.~\rf{Uone} and \rf{Uthree} respectively)
do not contain any constant matrices and could hence
{\em a priori} correspond to
topologically non-trivial configurations (i.e. which would allow for
fermion level crossing).
However, it is easy to see that these classes are empty.
It readily follows from the trivial fact that $\pi_2 (S^1)=0$.

\subsection{Spectrum of the Dirac Operator}
We now consider the Dirac equation
\beq
\sigma_i (\partial_i -iA_i(x)) \psi(x) =i\lambda
\psi(x)
\label{dirac}
\eeq
in the case of an external gauge field obeying \eq{AparG} with $S(x)\in
SU(2)\times U(1).$
We can show that there is a pairing up of non-zero levels of the Dirac
\op .
To that effect we make a parity reflection of the Dirac equation \eq{dirac}.
It is easy to see by using \eq{AparG} that
\beq
\sigma_i S(x)(\partial_i -iA_i(x)) (S^{-1}(x) \psi(-x)) =- i\lambda
\psi(-x) .
\label{diracP}
\eeq
Thus we see that the wave function
\beq
\psi^P (x) = S^{-1}(x) \psi (-x)
\eeq
corresponds to an eigenvalue $-\l .$
Therefore if one deforms an external gauge field among the fields
obeying \eq{AparG} for some $S(x)$ then the number of zero modes of the
Dirac \op \ is a topological invariant modulo 2.
Hence the fields obeying \eq{AparG} split into two classes:
in the first one the Dirac \op \ has an even number of zero
modes and in the second one it has an odd number of zero modes.
The first class is topologically trivial since it contains the vacuum
configuration.
The second one contains the sphaleron configuration and is non-trivial
in the sense that any configuration that belongs to it cannot be continuously
deformed into a vacuum \con \ as long as we keep \eq{AparG} satisfied.

Thus if we compare the present general cases to that considered in
Ref.\,\cite{topclass} (for the gauge field odd under the conventional
parity reflection) we can see that we still have two
topologically distinct classes while these classes are enlarged as
compared to those of Ref.\,\cite{topclass}.

This
 analysis actually gives the following simple interpretation of the above
classification of the gauge fields with definite generalized parity.
The subspace in the space of the gauge fields which corresponds to those with
definite generalized parity can be continuously constructed starting with the
gauge fields with definite ordinary parity.
Indeed the latter correspond to the case of the group element $S=1$ in
\eq{AparG}.
They are separated into two disconnected homotopy classes.
Let us now continuously deform the matrix $S$ away from 1.
By such a procedure we extend our homotopy classes to two disconnected
varieties, which are infinitely larger than the classes of gauge fields
with definite ordinary parity.
The fact that these two varieties are larger than those for $S=1$ was
actually
proven in section 3.1 above, where it was shown that there exist gauge field
configurations which are invariant under a generalized parity transformations
and are not gauge equivalent to those which are invariant under the ordinary
parity transformations.
Some explicit examples of such nontrivial deformations given in
section 5 support this statement.

Moreover as follows from the above analysis of the spectrum of the Dirac
operator
the number of zero modes of the Dirac operator remains a topological invariant
modulo 2.
That means that the two varieties which are constructed by deformations of the
group element $S$ remain homotopically disconnected.

An important question is now if the enlarged classes contain gauge
fields with the Chern-Simons numbers different from $0$ and $1/2$
(modulo an integer).
We address this question in next section.

\section{The Chern-Simons Functional}
\setcounter{equation}{0}

Let us first calculate the value of the Chern-Simons functional
for a gauge field obeying \eq{AparG}.
This functional is defined as follows
\beq
CS[A] = \fr 1{8\pi^2} \int_{D^3} {\rm Tr} (AdA-i\fr 23 A^3) .
\eeq
For odd-parity \con s $CS=0 .$
For the fields odd under the generalized parity we have
\beq
CS[A] = \fr 1{8\pi^2} \int_{D^3} {\rm Tr} (A(x) dA (x) -i\fr 23 A^3 (x)) =
\label{CSG1}
\eeq
$$=- \fr 1{8\pi^2} \int_{D^3} {\rm Tr} (A(-x)dA (-x) +i\fr 23 A^3 (-x))
=$$
$$=- \fr 1{8\pi^2} \int_{D^3} {\rm Tr} (A^S (x)dA^S (x) -i\fr 23 (A^S)^3
(x))= -CS[A^S] ,$$
where
\beq
A^S= S(x)(i\partial_i + A_i (x)) S^{-1}(x) .
\eeq
On the other hand we have the identity
\beq
CS[A^S]= CS[A] - S_{WZW} [S] ,
\label{iden}
\eeq
where
\beq
S_{WZW}[S] = \fr 1{24\pi^2} \int_{D^3} {\rm Tr} (dS S^{-1})^3
\eeq
and we assume that $S^{-1}dS \sim 1/|x|^{1+ \epsilon}$ at infinity
($\epsilon > 0$).
Substituting this identity (\ref{iden}) into \eq{CSG1} we get
\beq
CS[A]= \fr 12 S_{WZW}[S] .
\eeq
One can see that this value reduces to zero for any gauge field $A_i$
odd under the ordinary parity reflection since in this case $S=1 .$

However the value of the Chern-Simons functional (4.6) is not yet
associated with a given gauge field.
To define correctly the Chern-Simons number we have to make
a gauge transformation with an appropriate parameter $U^{\dag}$
in order to remove field at infinity, and then
evaluate the Chern-Simons functional for the gauge transformed
field $W = A^{U^+} .$
The field $W$ is assumed to decrease rapidly like $\sim 1/|x|^3$
at infinity.
Using \eq{iden} one can get
\beq
CS[W] = CS[A] + S_{WZW} [U] = \fr 12 S_{WZW} [S] + S_{WZW}[U] .
\label{CSW}
\eeq
For group elements $U(x)$ which are odd (even) at infinity,
whose corresponding gauge fields are odd under the ordinary parity
transformation, $S_{WZW}=1/2 \; (S_{WZW}=0)$ modulo an integer.
For gauge fields odd under the generalized parity both
$S_{WZW} [S]$ and $S_{WZW} [U]$ may be fractional.
Thus one could expect that the value of the Chern-Simons functional
can assume any fractional number when we enlarge the topologically
distinct classes. In what follows we will show that this is not the
case. First, observe that
the group elements $U$ and $S$ are related by the following \cn \
at infinity which can be deduced from \eq{AparG}
\beq
(dU U^{-1})(x) = - (d(SU)U^{-1}S^{-1})(-x) .
\label{bound}
\eeq
Let us calculate a variation of $CS[W]$ under a small deformation of
$S(x) .$
The variation of the Wess-Zumino-Witten functional is given by eq.(2.17).
Thus we see that the variation depends only on the behaviour of the
group element at infinity.
Changing $x\to -x$ under the integral and using \eq{bound} one can
easily get
\beq
\delta S_{WZW}[U] = -\delta S_{WZW}[U] -\delta S_{WZW}[S] ,
\eeq
and hence
\beq
\delta S_{WZW}[U] + \fr 12 \delta S_{WZW}[S] = 0.
\eeq
Thus we see that $CS[W]$ is invariant under continuous deformations of
the group element $S(x) .$
Therefore the only available values of the Chern-Simons functional
within our two topological classes are $0$ and $1/2$ modulo an integer.

We conclude that the space of gauge fields odd under a generalized
parity has a ${\bf Z}_2$ structure similar to the space of the odd-parity
fields.

In the previous section
we have shown that
the Wess-Zumino-Witten functional realizes
a homomorphism into ${\bf Z}_2$ from the set of the $SU(2)$ group valued
functions
corresponding to ordinary odd-parity gauge fields $A_i$. Such a homomorphism
maps a product of two $SU(2)$ group-valued functions
into a sum of the values of the Wess-Zumino-Witten functionals for each
of them.
In the case of a generalized parity we do not have such a homomorphism
because a product of two group elements corresponding to a pure-gauge
behaviour at infinity is not in general an element of the same sort.

\section{A Multitude of Sphaleron Deformations}
\setcounter{equation}{0}

We now proceed to give concrete examples which illustrate the
${\bf Z}_2$ structure in the space of gauge field
configurations. The purpose is to demonstrate that there is an
infinity of background configurations other than the sphaleron
which admit
an odd number of fermion zero modes and therefore contribute to
baryon-number changing processes in the hot electroweak theory.
The configurations must approach pure gauge at infinity and we take them
to be of the form
\begin{eqnarray}
A_i (x) &=&-i v(x) (\partial_i U (x)) U^{-1} (x) ,
\label{gauge}\\*[1ex]
\Phi(x)&=& f(x)\;U(x)\, \colvec,
\end{eqnarray}
where $v$ and $f$ are profile functions with the property that $v,f\to 1$ at
spatial infinity.

It is important to realize that the spectrum of the Dirac
equation (neglecting Yukawa couplings) depends only on the background gauge
fields,
and therefore the number of Dirac zero modes in a given gauge background
will be the same in the broken phase as in the hot symmetric phase of the
theory, in which the
Higgs-field expectation value is zero. We thus obtain a means of understanding
the mechanism of
baryon-number changing
transitions in the symmetric phase as well as in the broken phase. The effect
of Yukawa
couplings in the broken phase is merely to deform slightly the set of
configurations for which
fermionic zero modes occur (see Ref. \cite{ajn}) without reducing the number of
such configurations.

In the symmetric phase, the energy density of the (static) gauge configuration
\rf{gauge} is ${\rm Tr}\,F^2/(4 g^2)$, where
\beq
\label{fkl}
F_{kl} = -i (\d_{[k}v)(\d_{l]} U)U^{-1} + i v(1-v) (\d_{[k}U)(\d_{l]}U^{-1})
\eeq
and $A_{[k} B_{l]}\equiv A_k B_l - A_l B_k$. The energy density will be
singular
at points where the group element $U$ is multivalued. If the set of singular
points
constitutes a one-dimensional string, it is easy to show that integrability
of the energy density requires the
behavior $v\sim\rho^k$, $k>1$ of the profile function near the string, where
$\rho$ is the
perpendicular distance to the string. It follows that the gauge-field
configuration is,
also in the symmetric phase, that of an embedded Nielsen-Olesen vortex
\cite{embedded,ola}.
On the other hand, if the group element $U$ is multivalued at a
finite number of points, these point singularities may be suppressed simply by
providing
appropriate isolated zeros of the profile function $v$.

Let us apply our classification of gauge fields, developed in section 3,
to Nielsen-Olesen vortex-like
deformations of the sphaleron, using the circular string loop given by
\beq
A_i = -i v(\rho) \partial_i U U^{-1},
\eeq
where $\rho=\sqrt{(r-x_0)^2 + z^2}$ and $r,z$ are cylindrical coordinates. This
loop
has radius $x_0$ and resides in the plane $z=0$ with its center at $r=z=0$
(see Fig. 1).
\begin{figure}
\begin{center}
\begin{minipage}{11cm}
\begin{center}
\leavevmode
\epsfxsize=10cm \epsfbox{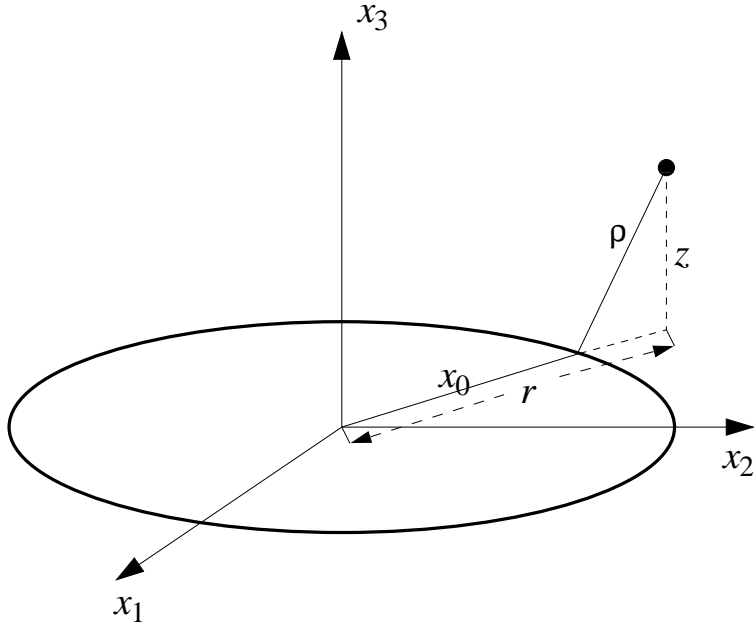}
\end{center}
\caption{{\sl Definition of cylindrical coordinates for circular
loop configurations.}}
\end{minipage}
\end{center}
\end{figure}
The profile function $v$ here satisfies $v(0)=0,$ $v(\infty)=1$, and its zeros
define the center of the vortex string.

Let us first consider loops of $W$-string \cite{Wstring,embedded}.
The $W$-string is a Nielsen-Olesen
vortex \cite{holger} embedded in the electroweak theory and is characterized by
an $SU(2)$ vector potential
which wraps around the string. Loops of this string have been studied
in Ref. \cite{topclass}. The simplest loop configuration corresponds to the
$SU(2)$ group element
\beq
\label{Wloop}
U_W= \frac{1}{\sqrt{(r-x_0)^2
+z^2}} \left( \begin{array}{cc} r-x_0 & z\\ -z & r-x_0 \end{array}
\right) .
\eeq
More generally, one can apply a ``twist'' which depends on the azimuthal angle
$\phi$ and
changes as one
goes along the loop. Consider first the group elements
\beq
U =
\exp (i(2n+1)\phi \tau_3/2)\! \frac{1}{\sqrt{(r-x_0)^2
+z^2}} \left(\! \begin{array}{cc} r-x_0 & z\\ -z & r-x_0 \end{array}\!
\right) \exp (i(2m+1)\phi\tau_3/2) ,
\label{mnloop}
\eeq
where $n$, $m$ are integers.
These group elements have a definite eigenvalue under the ordinary parity
transformation. We find that
for odd $m+n$ the matrix $U$ is even under parity, and the corresponding
configurations are topologically trivial (see section 2) with integer
Chern-Simons
number.
In contrast, the configurations with even $m+n$ correspond to odd $U$ matrices
and are topologically non-trivial with half-integer CS number. Therefore, the
configurations
given by \eq{mnloop}
for any $x_0$, $m$, $n$ and even $m+n$ admit an odd number of fermion zero
modes
and are as relevant as the sphaleron for baryon-number changing processes at
sufficiently
high temperature.

The topologically non-trivial configurations are deformations of the sphaleron,
as
can be seen from the following example. Consider the symmetrically
twisted loop with $m=n=0$ and take the limit
of a collapsed loop ($x_0 \to 0$). We obtain
\begin{eqnarray}
U&\to&
\exp (i\phi \tau_3/2)
\frac{1}{\sqrt{r^2
+z^2}} \left( \begin{array}{cc} r & z\\ -z & r \end{array}
\right)
\exp (i\phi\tau_3/2)\nonumber\\*
&=&\frac{1}{\sqrt{r^2 +z^2}} \left( \begin{array}{cc} x+iy & z\\ -z
& x-iy \end{array} \right) =-i\tau_2 U_{sph}\, i\tau_3\ ,
\end{eqnarray}
where $U_{sph}$ corresponds to the sphaleron configuration of Eq.~\rf{sphaU}.
Thus the collapsed loop coincides (up to a profile function) with the
sphaleron configuration.

For the particular values $n=0$, $m=-1$ one obtains an
asymmetrically twisted, topologically trivial, loop which can be
interpreted as a bound state of two sphalerons \cite{topclass}. In this case
the $SU(2)$
group element can be shown explicitly to be continuously connected to
$U=1$.

We  now turn to examples of gauge-field configurations with
a definite generalized parity (see section 3).
Consider the twisted loops corresponding to
\beq
U = \exp (im\phi \tau_3) \frac{1}{\sqrt{(r-x_0)^2 + z^2}}
\left( \begin{array}{cc} r-x_0 & z\\ -z & r-x_0 \end{array} \right)
\exp ( in \phi \tau_3) ,
\label{suloop}
\eeq
where $m$ and $n$ are
integers.
This matrix corresponds to a gauge field which is odd under a
generalized parity transformation
 and obviously belongs to class 2 in the classification of
section 3.
Hence the corresponding gauge field of the twisted
loop is topologically trivial.

In Ref. \cite{topclass} it was shown that the configuration given by
\eq{mnloop} for general
$n$ and $m$ can be
deformed, in a process referred to as ``splitting'',
into a configuration which corresponds to a collection of separated
simple fundamental ($m=n=0$ and $m=-1$, $n=0$) loops.
That analysis can be easily generalized to the $SU(2)\times U(1)$ gauge
theory and to gauge-field configurations which have a definite
generalized parity.

Let us now investigate the possibility of
$W$-string loops with an $SU(2)\times U(1)$ twist.
They would correspond to group elements $U_L^{mn}$ or
$U_R^{mn}$ given by
\beq
U_L^{mn} = \exp ( im \phi \tau_3)
\exp (ik\phi \frac{1+\tau_3}{2}) \frac{1}{\sqrt{(r-x_0)^2 +z^2}}
\left( \begin{array}{cc} r-x_0 & z\\ -z & r-x_0 \end{array} \right)
\exp ( in \phi \tau_3)
\label{uloopln}
\eeq
and
\beq
U_R^{mn} =\frac{1}{\sqrt{(r-x_0)^2 +z^2}} \exp ( im \phi \tau_3)
\left( \begin{array}{cc} r-x_0 & z\\ -z & r-x_0 \end{array} \right)
\exp (ik\phi \frac{1+\tau_3}{2}) \exp ( in \phi \tau_3) .
\label{ulooprn}
\eeq
where $k$, $m$, and $n$ are integers. These configurations have been
considered previously \cite{Wloops} in the special case $m=n=0$.

Although the matrices $U_L^{mn}$ and $U_R^{mn}$ fit into our classification of
gauge-field
configurations with definite generalized parity, we shall
show below that they correspond to configurations of infinite energy and will
be physically
allowed only if one introduces at least one additional vortex loop linked with
the original one.
This is because their twists as a function of the angle $\phi$ are singular on
the line
$r=0$. In fact, any $U(1)$ twist or a twist with a $U(1)$ component will
require such an
additional loop. In contrast, in the case of the $SU(2)$ gauge group the
singularity at $r=0$
can be easily eliminated without the need for extra loops.
To prove these statements we proceed as follows.

Consider a single closed vortex loop $C_1$ of any shape ($C_1$ being the set of
zeros of
$v$), subject to the condition
that it does not link with itself (i.e.\ is not knotted). This means that
$L(C_1,C_1)=0$, where
the linking number of two arbitrary curves $C_1$ and $C_2$ is defined by
\cite{yellowbook}
\beq
\label{link}
L(C_1,C_2)={{1}\over{4\pi}} \oint_{C_1}\oint_{C_2} {{
d{\bf x}_1\cdot d{\bf x}_2 \times ({\bf x}_1-{\bf x}_2)}\over{
{|{\bf x}_1-{\bf x}_2|}^3 }} .
\eeq
We first assume that the group element $U$ is single-valued everywhere except
on $C_1$.
Consider then a second closed curve $\gamma_1(\epsilon)$ following $C_1$ at a
distance
$\epsilon$ (see Fig. 2)
\begin{figure}
\begin{center}
\begin{minipage}{11cm}
\begin{center}
\leavevmode
\epsfxsize=10.5cm \epsfbox[131 380 436 580]{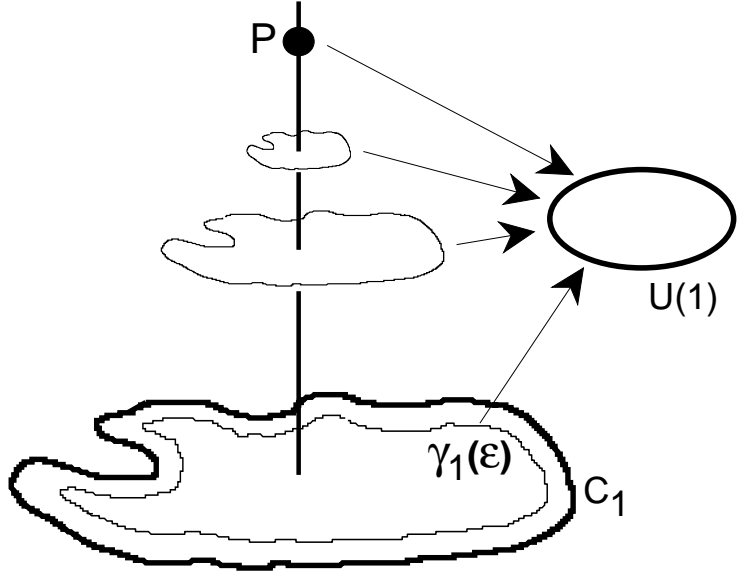}
\end{center}
\caption{{\sl A single closed vortex loop $C_1$ accompanied by a test path
$\gamma_1(\epsilon)$ tracing it at a distance $\epsilon$.
If the map $\gamma_1(\epsilon)\mapsto U(1)$ has non-zero winding number,
and if it is continuous as $\gamma_1(\epsilon)$ is shrunk to a point
{\rm P}, then the map is multivalued at {\rm P}.}}
\end{minipage}
\end{center}
\end{figure}
but not linking with it, i.e.\ $L(C_1,\gamma_1(\epsilon))=0$.
The assignment of
group elements $U$ to points on the curve $\gamma_1(\epsilon)$ is,
topologically,
a map from a circle $S^1$ into the group. If the group is $U(1)$ (which is
topologically equivalent
to $S^1$), such a map is characterized
by an integer topological winding number which measures the number of times $U$
runs around
$U(1)$ as one traverses the closed loop $\gamma_1(\epsilon)$. Since
$\gamma_1(\epsilon)$
does not link with $C_1$, it may be contracted within $R^3\setminus C_1$
to a point $P$ along any surface bounded by
$\gamma_1(\epsilon)$. If the winding number of the map is nonzero, the image of
the
map cannot be contracted within $U(1)$ and
the group element $U$ will be multivalued at $P$. By smoothly varying the
surface along which
$\gamma_1(\epsilon)$ is contracted, the set of points $P$ forms a
one-dimensional curve along
which $U$ is multi-valued.
The ensuing singularity must then be compensated by a zero in the profile
function $v$,
and we are presented with
another vortex
$C_2$ which passes through the loop $C_1$. Because we require pure gauge
behavior at spatial
infinity
in all directions, $C_2$ must be a closed loop which links with $C_1$. A
similar argument,
valid for $U(1)$ and configurations with $U(1)$ factors, was arrived at
independently in
Ref. \cite{dziarmaga}.

In the case of larger gauge groups, such as $SU(2)$ which is isomorphic to the
sphere $S^3$,
the image of the map from $\gamma_1(\epsilon)$ to the group will be a closed
curve embedded in
the simply connected group manifold $M$ of dimension $d>1$, and this curve can
be easily contracted
within $M$ to a unique element as $\gamma_1(\epsilon)$ is contracted to $P$.
Therefore, $U$ will
be single-valued everywhere except on $C_1$, and there is no need to introduce
another vortex loop.
The singularity in the $SU(2)$ configurations \rf{mnloop} and \rf{suloop}
arises
because their particular form requires  that $U$ take values in a
one-dimensional
subset of $M=SU(2)$ as $r\to 0$ for constant $z$.

Let us now give examples of regularized versions of the SU(2) configurations.
Consider first the simplest non-trivial loop
with the $U$ matrix given in \eq{mnloop} with $m=n=0.$
We want to write down an $SU(2)$ group element which
asymptotically coincides with $U$ far from the $z$-axis going through
the center of the loop perpendicular to its plane, and this
regularized group element is to be non-singular near this axis.
Such a group element is given by the following expression
\beq
U= \exp (i\phi \tau_3 /2)  U_0 \exp (i\phi \tau_3 /2) ,
\eeq
where
\beq
U_0 (x,y,z) = \frac{1}{\sqrt{(r-x_0)^2 \frac{r^2}{r^2 +
\alpha^2} +z^2}}
\left( \begin{array}{cc} (r-x_0)
\frac{r}{r+i\alpha} & z \\ -z  & (r-x_0)
\frac{r}{r-i\alpha} \end{array} \right)
\eeq
Here $\alpha$ is a small parameter.
This group element is odd under ordinary parity but
is singular at the point $r=z=0.$
The singularity can be suppressed by assuming an appropriate
zero of the profile function $v$ at this point.
In turn such a zero of the profile function does not correspond
to any string linked with the original loop and does not spoil the
pure gauge behaviour of the gauge field at infinity.

Let us now consider the $U$ matrix given in
\eq{mnloop} for the trivial loop configuration with $n=0$, $m=-1$.
The regularized version reads
\beq
U= \exp( i\phi \tau_3 /2) \tilde{U}_0
\exp (-i\phi \tau_3/2) ,
\eeq
where
\beq
\tilde{U}_0 =\frac{1}{\sqrt{(r-x_0)^2 +z^2 \frac{r^2}{r^2+\alpha^2}}}
\left( \begin{array}{cc} r-x_0 & z\frac{r}{r+i\alpha}\\
-z\frac{r}{r-i\alpha}  & r-x_0 \end{array} \right) .
\eeq

A similar modification can be used for general values of $m$ and $n$.
Regularized versions of Eqs.~\rf{mnloop} and \rf{suloop} can be obtained by
substitution of the above regularized group elements and by the
following substitution
\beq
e^{i\phi\tau_3} \to
e^{i\phi \tau_3 /2} (U_0)_{reg}
e^{i\phi \tau_3 /2} \cdot
e^{-i\phi \tau_3 /2} (U_0^{-1})_{reg}
e^{i\phi \tau_3 /2} ,
\eeq
where $(U_0)_{reg} =U_0$ and $(U_0^{-1})_{reg} (x,y,z) = \tilde{U}_0
(x,y,-z).$ Here
 each factor is regular as was seen above.
The resulting group element has a singularity at
$r=z=0$ which can be suppressed by introducing an appropriate zero of
the profile function. Because it is odd under ordinary parity, the regularized
versions of Eqs.~\rf{mnloop} and \rf{suloop}, obtained by the
above substitutions, have the
same parity properties as the unregularized ones.

Let us now consider an electro-weak $Z$-string segment with a
monopole and an anti-monopole ('t Hooft-Polyakov)
attached to its ends \cite{nambu}. The Z-string
\cite{Zstring} is an
embedded Nielsen-Olesen vortex, the core of which is a flux tube leading the
electroweak
$Z$ field from the monopole to the anti-monopole.
In our notation the gauge field for this configuration has the form \rf{gauge}
where $v$ is an appropriate profile function and
\beq
U(x) = \left( \begin{array}{cc} \cos (\Theta/2) & \sin (\Theta/2)
e^{-i\phi} \\ \sin (\Theta/2) e^{i\phi} & - \cos (\Theta/2) \end{array}
\right) .
\label{nambuU}
\eeq
Here the angle variable $\Theta\in [0,\pi]$ is defined as follows
\beq
\cos \Theta =\cos \theta_m -\cos \theta_{\bar{m}} +1 .
\label{Theta}
\eeq
The polar angles $\theta_m$ and $\theta_{\bar{m}}$ are defined in
Fig. 3.
\begin{figure}
\begin{center}
\begin{minipage}{11cm}
\begin{center}
\leavevmode
\epsfysize=6cm \epsfbox{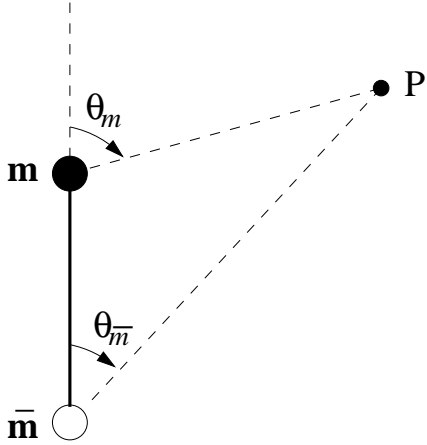}
\end{center}
\caption{{\sl Definition of polar angles in the
Nambu monopole-antimonopole configuration.}}
\end{minipage}
\end{center}
\end{figure}
It can be seen that $\Theta$ is approximately the angle subtended by the string
segment between the two monopoles at the position $x$.
The $U$ matrix in \eq{nambuU} corresponds to the Higgs field configuration
\beq
\Phi_{m\bar{m}} = f(x)\;U \,\colvec =
f(x) \left( \begin{array}{c} \cos (\Theta/2) \\
\sin (\Theta/2) e^{i\phi} \end{array} \right) ,
\eeq
where $f(x)$ is a profile function.

One can check that the configuration given by Eqs.~\rf{nambuU} and \rf{Theta},
in the limit $\theta_{\bar{m}} \to 0$, approaches the monopole configuration
\beq
U_{mon} = U_1 (\theta_m ,\phi) \equiv
\left( \begin{array}{cc} \cos (\theta_m/2) & \sin
(\theta_m/2)
e^{-i\phi} \\ \sin (\theta_m/2) e^{i\phi} & - \cos (\theta_m/2) \end{array}
\right),
\label{mono}
\eeq
and, similarly, that it approaches the anti-monopole configuration given by
\beq
U_{antimon} = U_1 (\frac{\pi}{2} - \theta_{\bar{m}},\phi) =
\left( \begin{array}{cc} \sin (\theta_{\bar{m}}/2)
& \cos (\theta_{\bar{m}}/2)
e^{-i\phi} \\ \cos (\theta_{\bar{m}}/2) e^{i\phi} & - \sin
(\theta_{\bar{m}}/2) \end{array}
\right).
\label{antimono}
\eeq
in the limit $\theta_m \to \pi$.
The string singularity is localized on the straight line joining the
monopole and anti-monopole and corresponds to $\theta_{\bar{m}}= 0$
and $\theta_m = \pi .$
The angle $\phi$ is measured around this axis.

These $U$ matrices correspond to the following Higgs fields near the
monopole
\beq
\Phi_{mon} = f(x) U_1 (\theta_m ,\phi) \colvec=
f(x) \left( \begin{array}{c} \cos (\theta_m /2) \\
\sin (\theta_m /2) e^{i\phi} \end{array} \right) ,
\eeq
and near the anti-monopole
\beq
\Phi_{antimon} = f(x) U_1 (\frac{\pi}{2} - \theta_{\bar{m}} ,\phi) \colvec=
f(x) \left( \begin{array}{c} \sin (\theta_{\bar{m}} /2) \\
\cos (\theta_{\bar{m}} /2) e^{i\phi} \end{array} \right) .
\eeq
The parity transformation corresponds to
\beq
\phi \to \phi +\pi,\;\;\; \theta_m \to \pi -\theta_{\bar{m}} ,
\;\;\; \theta_{\bar{m}} \to \pi-\theta_m .
\label{parity}
\eeq
It is easy to see that the matrix $U$ defined by Eq.~\rf{nambuU}
satisfies
\beq
U(-x) = -i\tau_3 U(x) i\tau_3 ,
\eeq
and hence belongs to the trivial class 2 of our classification.
Thus this configuration has Chern-Simons number 0 and does not permit
fermionic level crossing.
Note that this conclusion agrees with the analysis of
Ref. \cite{hind}.

Let us consider instead a ``twisted'' configuration, introduced in
Ref.\ \cite{Ztwist},
of a monopole and
an anti-monopole joined by a segment of electroweak string.
The gauge field is given by the same formula \rf{gauge} with
\beq
U_{\gamma} (\theta_m ,\theta_{\bar{m}} ,\phi) =
\left( \begin{array}{cc} a & -b^{\ast} \\ b &  a^{\ast} \end{array} \right) ,
\label{fstring2}
\eeq
where
\beq
a= \sin (\theta_m/2) \sin (\theta_{\bar{m}}/2) e^{i\gamma}
+\cos (\theta_m/2) \cos (\theta_{\bar{m}}/2) ,
\eeq
\beq
b= \sin (\theta_m/2) \cos (\theta_{\bar{m}}/2) e^{i\phi} -
\cos (\theta_m/2) \sin (\theta_{\bar{m}}/2) e^{i(\phi -\gamma)} .
\eeq
Here the constant $\gamma$ is the angle of twist.
This matrix $U_\gamma$ corresponds to the following Higgs field
\beq
\Phi_{m\bar{m}} =f(x) \left(\begin{array}{c} a \\ b \end{array}\right).
\eeq
It is easy to see that as $\theta_{\bar{m}} \to 0$ we get the monopole
configuration with the $U$ matrix given by Eq.\rf{mono} and as $\theta_m
\to \pi$ one has the anti-monopole field with $U$ defined in
Eq.\rf{antimono}, provided we perform the rotation $\phi \to \phi +\gamma.$
The configuration of Eq.\rf{fstring2} also has the usual
string singularity along the line that joins the monopoles.
Indeed let us take $\theta_{\bar{m}}=0$ and $\theta_m=\pi .$
In the domain near this
line it is easy to see that the Higgs field reduces to
\beq
\Phi_{m\bar{m}} =\left( \begin{array}{c} 0 \\ e^{i\phi}
\end{array} \right) f(x).
\eeq

Let us now consider the parity properties of the configuration given by
Eqs.~\rf{gauge} and \rf{fstring2}.
It is easy to check that this configuration has a definite generalized
parity only if $\gamma =0$ or $\pi .$
At $\gamma=0$ the matrix $U_\gamma$ is even under the ordinary parity
transformation
\rf{parity}.
Hence it corresponds to a topologically trivial field
configuration connected to a vacuum; it has Chern-Simons number
0 and does not permit fermionic level crossing.
In contrast, for $\gamma=\pi$ the group-valued function $U_\gamma$ is odd under
parity.
The corresponding configuration is topologically non-trivial and continuously
related to a
sphaleron field.
It can be seen easily that, when the string joining the monopole
and anti-monopole is collapsed to a point, the angles $\theta_m$ and
$\theta_{\bar{m}}$ can be identified and we obtain the sphaleron
configuration given in Eq.~\rf{sphaU}, up to a factor $\tau_3$.
Hence the configuration has Chern-Simons number 1/2 and is
relevant for rapid fermion number violation in the hot theory.

When we continuously change the angle of twist $\gamma$ from 0 to
$\pi$ the configuration goes through fields which do not have
any definite (generalized) parity eigenvalue.
Therefore we cannot control the Chern-Simons number of the field under
such a deformation.
However we see that the Chern-Simons number changes continuously
from 0 to 1/2 .

Let us now consider deformations, with definite generalized parity,
of the above monopole-antimonopole
configurations (with $\gamma =0,\pi$).
Such deformations can be constructed easily if we observe that
for the trivial configuration ($\gamma =0$)
\beq
U_{\gamma=0} (\theta_m ,\theta_{\bar{m}} ,\phi)
=U_1 (\theta_m ,\phi) U_1 (\theta_{\bar{m}} ,\phi) ,
\label{trivspl}
\eeq
while for the non-trivial one ($\gamma =\pi$)
\beq
U_{\gamma=\pi} (\theta_m ,\theta_{\bar{m}} ,\phi)
=U_1 (\theta_m ,\phi) \sigma_3 U_1 (\theta_{\bar{m}} ,\phi) \sigma_3 ,
\label{nonspl}
\eeq
where $U_1$ is defined in \eq{mono}. Following the approach of Ref.
\cite{topclass},
we shall see below that the monopole-antimonopole configurations may be
continuously deformed
into two segments of electroweak string, each segment joining a monopole with
an anti-monopole.

Indeed let us consider the topologically trivial configuration with the
$U_{\gamma=0}$ matrix.
Define $\theta_{\bar{m}'}$ and $\theta_{m'}$ to be the polar angles measured
from
two points $\bar{m}'$ and $m'$ located on the line between the initial monopole
and
anti-monopole (see Fig. 4).
\begin{figure}
\begin{center}
\begin{minipage}{11cm}
\begin{center}
\leavevmode
\epsfysize=6cm \epsfbox{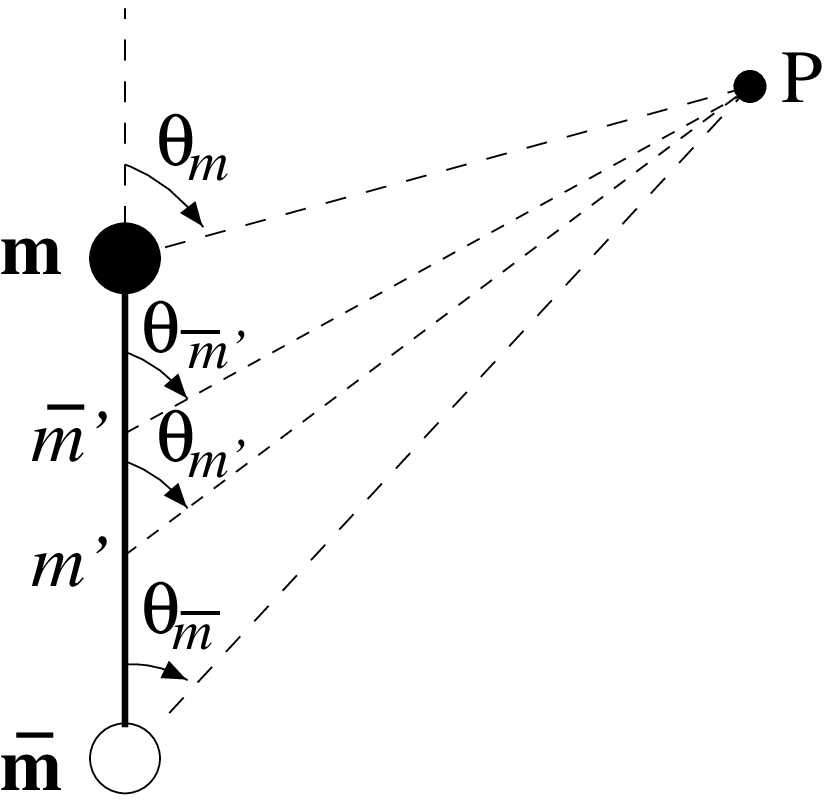}
\end{center}
\caption{{\sl The splitting of a $Z$-string segment, joining a
monopole-antimonopole
pair, into two $Z$-string segments by the production of a second
monopole-antimonopole pair.}}
\end{minipage}
\end{center}
\end{figure}
When both $\bar{m}'$ and $m'$ coincide with the midpoint
of this line, so that $\theta_{\bar{m}'}=\theta_{m'}$, we have
\beq
U_{\gamma=0} (\theta_m ,\theta_{\bar{m}} ,\phi) =
U_{\gamma=0} (\theta_m ,\theta_{\bar{m}'} ,\phi)
\times U_{\gamma=0}(\theta_{m'},\theta_{\bar{m}},\phi) ,
\label{u0split}
\eeq
This follows from \eq{trivspl} and the identity $U_1^2=1.$
Let us now deform the configuration continuously by moving the points
$\bar{m}'$ and $m'$ an equal distance in opposite directions along the axis of
the string,
so that
$\theta_{\bar{m}'}$ and $\theta_{m'}$ are no longer equal.
The points $\bar{m}'$ and $m'$ are associated with an anti-monopole and a
monopole
which emerge at the midpoint and then separate. As a result
the original string segment is
divided into two segments;
one between $m$ and $\bar{m}'$, the other between $m'$ and $\bar{m}$ (see Fig.
4).
Such a splitting is easily arranged by a deformation of the profile
function. In \eq{u0split} the multiplication sign $\times$ (which is ordinary
multiplication)
separates the factors corresponding to the field configurations
of the individual segments.
 The new configuration is even under a generalized parity
transformation with respect to the midpoint between the two string segments.
We have thus split the trivial configuration into two also
trivial ones.

One can also split the trivial configuration into two non-trivial
ones as follows
\begin{eqnarray}
U_{\gamma=0} &=&
U_1 (\theta_m ,\phi) \sigma_3 U_1 ( \theta_{\bar{m}'},\phi)
\sigma_3 \times
\sigma_3 \left( U_1 (\theta_{m'},\phi) \sigma_3
U_1 (\theta_{\bar{m}} ,\phi) \sigma_3 \right) \sigma_3 \nonumber\\*
&=&U_{\gamma=\pi} (\theta_m ,\theta_{\bar{m}'} ,\phi)
\times \sigma_3 U_{\gamma=\pi}(\theta_{m'},\theta_{\bar{m}},\phi) \sigma_3 .
\end{eqnarray}
Again the equality holds when $\theta_{\bar{m}'}=\theta_{m'}$ by virtue of
Eqs.~\rf{trivspl},
\rf{nonspl} and
$U_1^2=1$.
As before, we split a single string segment into two by moving ${\bar{m}'}$
and $m'$
in opposite directions from the midpoint of the string and  deforming
the profile
function accordingly.
The new configuration is obviously even under a generalized parity
transformation
with respect to the midpoint between the two segments and is
topologically trivial, while
each segment corresponds to a topologically non-trivial configuration.

This particular example shows clearly that there exist configurations which
globally admit
only an even number of zero modes of the Dirac equation, but consist of
localized excitations
each corresponding to a saddle-point-like configuration with an odd
number of zero modes
which can mediate
$B+L$ violating transitions. This illustrates the concept of
a fluid composed of localized $B+L$ violating excitations that was alluded to
in the introduction. The properties of such a ``fluid'' require further
investigation.
We shall return to this issue in the next section.

Similarly we can split the nontrivial configuration given by $U_{\gamma=\pi}$
into one non-trivial
and one trivial configuration along the axis of the string as follows.
\beq
U_1 (\theta_m ,\phi) \sigma_3 U_1 (\theta_{\bar{m}} ,\phi)
\sigma_3 =
U_1 (\theta_m ,\phi) U_1 ( \theta_{\bar{m}'},\phi)\times
U_1 (\theta_{m'},\phi) \sigma_3 U_1 (\theta_{\bar{m}} ,\phi)
\sigma_3 .
\eeq
We separate ${\bar{m}'}$ and $m'$ along the string axis
and get a new configuration, odd under a generalized parity transformation
with respect
to the midpoint between the two string segments.

We have thus described a set of deformations, with definite parity,
of the monopole-antimonopole configuration.
This analysis, together with that of Ref. \cite{topclass}
for the loop configurations, provides a list of
parity-preserving deformations of the string loops and string segments
at the classical level. This list includes an infinite set of
odd-parity deformations of the sphaleron which, like the sphaleron,
mediate baryon-number
changing processes near and above the electroweak phase transition
temperature. The analysis of the processes of splitting can be
instructive for understanding the dynamics of string loops and
segments at the quantum level.

We conclude this section by observing that the {\it new\/} electroweak
``sphaleron'' $S^*$  fits into our classification of
gauge fields. Its gauge-field configuration is defined by \eq{gauge} where
the matrix $U$, extracted from Ref. \cite{kl} with some amount of algebra,
is given by
\beq
\label{newS}
U = i\; {{
2 z x_i \tau_i - \left[ r^2 + z^2 + {({{d}\over{2}})}^2 \right] \tau_3}\over
{
\sqrt{
\left(r^2 + {(z-{{d}\over{2}})}^2\right)\!\left(r^2 +
{(z+{{d}\over{2}})}^2\right)}
}} .
\eeq
The profile function $v=v(r,z)$ satisfies $v(r,z)=v(r,-z)$ and has two zeros at
the positions $r=0$, $z=\pm d/2$. These zeros can be thought of as the loci of
a sphaleron $S$ and an anti-sphaleron $\bar{S}$ which together form the
bound state $S^*$.
The gauge fields constructed from $U$ and \eq{gauge}
are obviously odd under ordinary parity. Because $U$ is even, the configuration
$S^*$ is topologically trivial and plays no role in $B+L$ violating processes.

Similar analysis can be done for the multisphaleron
configurations with $CS= n/2$, $n \in {\bf Z}$. For the axisymmetric
ansatz given in \cite{Kleihaus}, multisphalerons with
$n$ odd have odd $U$ at infinity, possess an odd number of fermion
zero modes and
thus mediate $B+L$ violation. Because each class of the ${\bf Z}_2$
homotopy group is non-empty, we expect the existence of even-$n$
($CS\in {\bf Z}$) multisphaleron
solutions with (generalized)-odd gauge fields,
even $U$ at infinity, possessing an
even number of fermion zero modes. Such multisphalerons are connected to
the $S^*$ new sphaleron and the trivial vacua and do not mediate $B+L$
violation in the early universe.

\section{Conclusions and Discussion}
\setcounter{equation}{0}

In the present work we have addressed the issue of the topological origin
of $B+L$ violation in the electroweak theory.
The existence of a homotopy
group $[{\rm R}P^2 ,G/{\bf Z}_2]={\bf Z}_2$
for $G=SU(2N)$, $SO(2N)$ and $E_7$ implies
the existence of a ${\bf Z}_2$ structure in the gauge
sector of the electroweak theory in its symmetric phase as well as its
broken phase.
Because of this ${\bf Z}_2$ structure
the Chern-Simons number ($CS$) plays the new role of
a topological index for gauge fields which are odd under a generalized parity
transformation. For configurations with even-parity pure-gauge behaviour
at infinity, $CS$ takes an integer value $n$,
while for an odd-parity pure-gauge behaviour, $CS$ takes a value $n+1/2$.
Thus group theory implies the existence of two topologically distinct
infinite sets
of odd-parity gauge field configurations with a continuous range of energies.
This structure characterizes both the symmetric and the broken phase
of the theory and does not depend on the particular Higgs sector of the model.
In the broken phase, it
automatically accounts for the existence of a sphaleron
saddle point with energy  $E_{sph}\approx M_W/\alpha$ and Chern-Simons number
$CS=1/2$.

The important role of sphalerons and their deformations in $B+L$ violating
processes
is a
direct consequence of the topological
invariance modulo 2 of the number of zero modes of the Dirac operator in
their background.
This invariance implies that $B+L$ violating fermion level crossings occur
in the presence of {\it any\/} finite-energy (generalized) odd-parity gauge
field configuration
with an odd pure-gauge behaviour
at infinity ($CS=n+1/2$).
Odd-parity gauge field configurations with an
even pure-gauge behaviour at infinity, such as the new
$S^*$ ``sphaleron'',
admit an even number of fermion zero modes and do not mediate
$B+L$ violation.

We remark that, while the ${\bf Z}_2$ topological structure provides
us with an infinity of
gauge-Higgs field configurations with $B+L$ violating properties,
we cannot claim to have exhausted the whole space of such configurations. In
fact,
consider a path leading from the vacuum with $CS=n$ to the vacuum $CS=n+1$
over some odd-parity configuration with $CS=n+1/2$. It is easy to show from
continuity
of the Dirac spectrum that if a small even-parity gauge field $A_E$ is added to
every configuration
along the path, satisfying $A_E(n)=A_E(n+1)=0$, then an odd number of zero
modes must occur
for some Chern-Simons number $n<CS<n+1$.

Despite these findings, we assert that the space of generalized
odd-parity gauge fields with odd pure-gauge behavior at infinity constitutes
the backbone in the body of all configurations that mediate $B+L$ violation.
This is quite transparent in the broken phase of the theory ($T < E_{sph}$),
where the sphaleron saddle point {\it alone\/} governs the dynamics of thermal
transitions
between vacua.

As the temperature is raised, other configurations will begin to contribute to
the
thermal transition rate. Those with odd generalized parity are the first ones
to
become excited. This result comes about from studying configurations in a
neighborhood
of the sphaleron. More precisely, we consider the space $E$ of deformations of
the
sphaleron which are orthogonal to the unstable mode. Let us introduce an ${\cal
L}^2$
metric on $E$ and consider a sphere $S_\epsilon \in E$ with radius $\epsilon$
centered
at the origin (which corresponds to the sphaleron). On this compact sphere,
the energy functional assumes maximal and minimal values.
Because the energy functional is invariant under the generalized parity
transformation
defined in section 3.1, fields $A_i$ which are invariant under this discrete
symmetry
correspond to an extremum on $S_\epsilon$. It is natural to expect, by
continuity arguments,
that the extremum constitutes a minimum. This claim can be justified through
direct computation.

As the result of such an argument, on the sphere $S_\epsilon$ sphaleron
deformations with
odd generalized parity have lower energy than the rest. Therefore, as the
temperature is
raised, these sphaleron deformations are the first to be excited. The concept
that
generalized odd-parity configurations with odd pure-gauge behavior at infinity
are the
backbone of the configurations
mediating $B+L$ violation is thereby justified, at least for temperatures near
and not too
high above the sphaleron energy $E_{sph}$.

It is appropriate at this point to remark about the domain of
validity
of our arguments. In an $SU(2)$ gauge theory it was shown \cite{yaffe}
that for
$M_H > 10 M_W$ there emerges a doublet of saddle-point configurations of
lower energy than the $CS=1/2$ sphaleron with
$CS= 1/2 \pm \epsilon(M_H)$. For such a range of Higgs mass, and at
temperatures comparable to the energy of the saddle points, we would
expect the $B+L$ violating transitions to be dominated instead by gauge-field
configurations with $1/2-\epsilon < CS < 1/2$ and $1/2 < CS < 1/2 + \epsilon$.

While the density of $B+L$ violating sphaleron-like configurations is
Boltzmann suppressed for $T<E_{sph}$, the
expected monotonic increase of density of such configurations
as a function of temperature would provide an explanation for the unsuppressed
$B+L$ violating
transition rate in the symmetric phase ($\Gamma\propto T^4$), which cannot be
understood from
perturbative expansions around the sphaleron \cite{peter,dyakonov}.

We may further speculate that the large density of finite-energy configurations
homotopically equivalent to
the sphaleron will induce dynamically a $B+L$ violating quark-lepton condensate
in the symmetric phase.
The analogous phenomenon, in the presence of a nonzero density $\rho(0)$ of
 instanton energy levels near zero energy, has already been studied in the
context of chiral
symmetry breaking in QCD \cite{condensate}, where the quark condensate
$\langle \overline{\psi}\,\psi\rangle$ is proportional to $\rho(0)$.
In our case the magnitude of
the condensate would be proportional to the density of energy levels
of the three-dimensional \dop{} near zero energy.
At high temperature the condensate has to be proportional to a power of the
temperature since this is the only dimensionful parameter.

The form of such a fermionic condensate is restricted by the gauge
symmetry $SU(3)_c \times [SU(2) \times U(1)]_{EW}$ of the standard model.
The simplest form is
\beq
\langle\;\sum_{ijk} \epsilon_{ijk} (q_L^i q_L^j)\,( q_L^k l_L)\rangle,
\eeq
where $q_L$ and $l_L$ are the quark and lepton left-handed doublets and
the indices $i,j,k$ stand for the fundamental
representation of the $SU(3)_c$ group.
It is easy to see that the above expression has zero hypercharge
and hence is gauge invariant.
Since we have no direct experimental information on whether color symmetry
could have been spontaneously broken at high temperature, we should not exclude
color-charged
condensates from consideration.
 The possible existence of $B+L$ violating condensates in the symmetric phase
will be the subject of a future investigation.

A brief summary of the results of this paper was given in Ref.\,\cite{EPS}.

\section{Acknowledgments}
\setcounter{equation}{0}

A.\,J. is grateful to The Niels Bohr Institute for warm hospitality.
His research is partially supported by a NATO grant GRG 930395.
O.\,T. is grateful to R.\,Pisarski and the Brookhaven
National Laboratory Summer Programs for partial support and hospitality.
M.A. acknowledges partial support by Danmarks Grundforskningsfond
through its contribution to the establishment of the Theoretical Astrophysics
Center, as well as support from the European Union under contract number
ERBCHRXCT-940621.

\typeout{
****************************************************************}
\typeout{
* NOTE: YOU MAY NEED TO LATEX 3 TIMES TO GET THE FIGURES RIGHT *}
\typeout{
****************************************************************}
\end{document}